\documentclass[prb,showpacs,nobibnotes,nofootinbib]{revtex4}
\usepackage{amsmath}
\usepackage{amssymb}
\newcommand{\Tr}{{\rm Tr~}}
\newcommand{\cp}[1]{\ensuremath{\mathbb{CP}^{#1}}}
\newcommand{\mnras}{Mon. Not. Royal Astron. Soc.\ }

\begin{document}

\title{Magnification relations in gravitational lensing via multidimensional 
residue integrals}

\author{Neal Dalal}
\email{endall@astrophys.ucsd.edu}
\affiliation{Physics Department, University of California, San Diego, CA 92093}
\author{Jeffrey M. Rabin}
\email{jrabin@euclid.ucsd.edu}
\affiliation{Mathematics Department, University of California, San Diego, CA 92093}

\date{\today}

\begin{abstract}
We investigate the so-called magnification relations of gravitational
lensing models.  We show that multidimensional residue integrals
provide a simple explanation for the existence of these relations, and
an effective method of computation.  We illustrate the
method with several examples, thereby deriving new magnification
relations for galaxy lens models and microlensing (point mass lensing).
\end{abstract}

\pacs{02.10.Rn, 02.30.Fn, 98.62.Sb}

\maketitle

\renewcommand{\thefootnote}{\fnsymbol{footnote}}

\section{Introduction}
\label{sec:intro}
Gravitational lensing has proven to be not only important
astrophysically, but intriguing mathematically as well.  Mathematical
investigations of gravitational lens theory have yielded important
results and insights,\cite{bn86,sef,witt90,petters-morse,petters,lpw93}  
employing techniques and results from such disparate areas
as catastrophe theory, differential geometry and Morse theory.  In this
paper, we illustrate how another seemingly unrelated subject,
multidimensional residue calculus, applies to gravitational lensing,
and specifically we explain the origin of certain ``magnification relations''
that have been discussed in the lensing literature.\cite{wm95,rhie,dalal,wm99}
We additionally demonstrate that calculations of these magnification 
relations are enormously simplified using residue techniques, and 
illustrate the method by deriving several new results.

This paper is organized as follows.  In the remainder of this section,
we introduce the relevant terminology of gravitational lensing, and
describe the magnification relations.  In Section \ref{sec:trace}, we
express the problem in terms of residue calculations at the image
positions, and thereby relate it to a residue at infinity.  Using
one-dimensional residue calculus, we derive trace formulas for the
magnification relations for a subset of lens models.  In Section
\ref{sec:residue}, we consider the general class of lens models, and
describe how to perform the necessary multidimensional residue
integrals.  In Section \ref{sec:example}, we apply this formalism and
derive previously known results, as well as new results.  In Section
\ref{sec:discuss}, we summarize our results and discuss implications.
The material discussed in sections \ref{sec:trace}-\ref{sec:example}
may be unfamiliar to astronomers, and so we provide a simple procedure
which may be applied to models to obtain their magnification
relations, without requiring a detailed understanding of the
underlying mathematics.

\subsection{Gravitational lensing terminology}

Numerous excellent introductions to gravitational lensing have been
written, e.g. Refs.[\onlinecite{bn86,sef}]; here we briefly summarize some 
of the results and terminology relevant to our discussion.  The effects of
gravitational lensing can perhaps best be understood by considering
the time delay of trajectories connecting the lensed source to the
observer.\cite{bn86}  The time delay is a simple sum of two terms, a
``geometrical'' piece, and a ``gravitational'' piece.  Let 
$\phi_{\rm N}$ be the 3-D Newtonian potential, and 
$\psi=(2/Dc^2)\int\phi_{\rm N}\, dl$ the projected 2-D potential, 
where $D$ is a function of cosmology and the source and lens redshifts.  
Just as $\nabla^2\phi_{\rm N}=4\pi G\rho$, similarly
$\nabla_\perp^2\psi\propto4\pi G\Sigma$, where $\Sigma=\int\rho\, dl$ is
the surface density.  In the thin-screen, small deflection limit, the
time delay can be written as
\begin{equation}
\tau=\tau_0\left({1\over 2}|{\vec\Theta}-{\vec\beta}|^2-\psi\right),
\end{equation}
where the normalization $\tau_0$ depends upon cosmology and the lens
and source redshifts, ${\vec\Theta}$ is the image location on the sky,
and ${\vec\beta}$ is the location the source would have had, had not
the lens intervened.  Light rays follow null geodesics, which can be
shown to obey Fermat's principle.\cite{sef} Plotting time delay as a
function of the two angular coordinates on the sky, Fermat's principle
demands that images arise at the stationary points of this time delay
surface.  Setting the gradient of the time delay to zero gives the
so-called ``lens equations'', the (real) solutions of which are the
image positions.  Since the time delay surface can have multiple
stationary points, multiple images of a single source can arise, and
the image multiplicity can depend on the source position relative to
the lens.  The curves on the source plane separating regions of
different image multiplicity are called caustics.  As a source crosses
over a caustic, its image multiplicity changes by two, as a pair of
images either merge together and annihilate, or are created and move
apart.  The magnification $\mu$, which relates differential area
elements of the unlensed source to area elements of the lensed images,
is simply the inverse of the Jacobian $J$ of the mapping from
image coordinates ${\vec\Theta}$ to source coordinates ${\vec\beta}$.
Since the orientation of an image can be inverted relative to the
unlensed source, the magnification can have either sign.  Much of the
lensing literature adopts the convention of positive magnifications
(i.e. defining $\mu=1/|J|$); in this paper we always take the 
magnification to be signed, i.e. $\mu=1/J$.

\subsection{Lens models}

Two astrophysically important types of lenses are compact objects
(like stars or MACHOs) and galaxies.  The former class are
effectively point masses, and so have lensing potentials proportional
to the Green's function for the 2-D Laplacian, i.e. $\psi=m\log r$ 
where $m$ is proportional to the mass of the point lens.  Only the
weak-field regime is observationally relevant, so the potentials
linearly superpose for multiple point masses (as long as they are not
appreciably separated along the line of sight).  Galaxies have more 
complicated, extended mass distributions with correspondingly 
complicated lens potentials.  In principle, one may 
decompose the potential into eigenfunctions of the 2-D Laplacian\cite{koch91}
\begin{equation}
\psi=\sum_{m,n}(a_{mn}\cos m\theta +b_{mn}\sin m\theta)r^n.
\end{equation}
This has the advantage that each term in the expansion relates to a
corresponding multipole in the expansion of the surface density.  In
practice, it is necessary to truncate the series due to the limited
observational constraints.  Since galaxies are believed to have
roughly ``isothermal'' $\rho\sim r^{-2}$ profiles, such truncated
series generally consist of variations on the singular isothermal
sphere (SIS) potential $\psi=br$.  Two examples considered by
\textcite{koch91} are the SIS + elliptical potential 
$\psi=br(1+\gamma\cos 2\theta)$ and the SIS + external shear
$\psi=br+{\gamma\over 2}r^2\cos 2\theta$.  Another variation\cite{bk87}
is the singular isothermal ellipse (SIE) $\psi=bR=b\sqrt{x^2+y^2/q^2}$ 
with axis ratio $q$.  Other, more 
elaborate and more physically justified models have been employed in
the lensing literature; here we will focus on simple models such as
the above to avoid obfuscating the general method with heavy algebra.
In table \ref{tab1} we list the models considered in this paper.

Given a model for the lens, the lensed images of a given source may be 
found by solving the lens equations as described above.  For simple
potentials, the solutions are often analytic.  For example, consider
the the SIS + external shear model.  The stationarity equations become
\begin{eqnarray}
{{\partial\tau}\over{\partial r}}&=&r-s\cos(\theta-\theta_s)-b-\gamma r
\cos 2\theta=0,\\
{1\over r}{{\partial\tau}\over{\partial\theta}}&=&s\sin(\theta-\theta_s)
+\gamma r\sin 2\theta=0,
\end{eqnarray}
where $(r,\theta)$ is the image position in polar coordinates,
and $(s,\theta_s)$ is the source position.  From this, it is easy to
show that the quantity $u=e^{i\theta}$ satisfies the fourth degree
polynomial equation
\begin{equation} \label{extshear}
\gamma bu^4+(\gamma sv + {s\over v})u^3 - (sv+{{\gamma s}\over v})u
+\gamma b=0,
\end{equation}
with $v=e^{i\theta_s}$.  As a quartic equation in $u$, this can be solved
analytically, and from $u$ the image
coordinates $(r,\theta)$ follow simply.  For many models, it is
possible to eliminate all but one variable in a similar fashion and
thereby obtain analytic solutions, but for more realistic models,
the lens equations must be solved numerically.

\subsection{Magnification relations}

As mentioned above, there are in general multiple lensed images for a
given source.  These images lie at the discrete positions satisfying
the lens equations for a given source position, and the images have
different magnifications and orientations.  If we restrict ourselves
to purely real solutions of the lens equations (i.e. the physically
observable images) then the number of images changes when the source
crosses over a caustic.  The number of solutions, of course, does not
change, but instead a pair of complex solutions become real (for a
source crossing into a caustic) or a real pair become complex (for a
source crossing out of a caustic).\cite{petters} For certain lens
models, such as the simple potentials described above, there
may exist certain parameter ranges such
that {\sl all} solutions to the lens equations are real.  In such
cases, it has been shown that there exist interesting and surprising
relations among the image positions and magnifications.  First,
\textcite{wm95} considered lensing by a binary microlensing system,
involving two point masses, and derived the following result : when
the image multiplicity is maximized, the sum of the signed
magnifications of all the images is always 1.  That is,
$\sum_i\mu_i=1$, independent of quantities such as the lens masses,
separation, or source position (as long as the source is inside a
caustic).  This is quite an astonishing result, as the individual
image magnifications can vary wildly as the source position changes,
and even diverge as the source crosses a caustic.  Witt \& Mao derived
this result by using resultants to obtain a monic polynomial equation
satisfied by $\mu$, and noted that the sum of the roots is given by
the subleading coefficient.  \textcite{rhie}, using similar reasoning, 
showed that a similar result is true for an arbitrary number of
point mass lenses.  \textcite{dalal} extended this work to galaxy 
potentials like those described above.  Again, the magnification 
relations were obtained by using elimination theory (e.g. Gr\"obner
bases) to obtain monic polynomial equations in $\mu$; the ``total
magnification'' was then given by the subleading coefficient.  
We summarize these results in table \ref{tab1}.

Subsequently, \textcite{wm99} showed that for a particular class of
power-law models, there exist additional magnification relations,
involving not only $\mu$ but the image positions as well.  They
derived this result by separating the coupled lens equations into
disjoint $x$ and $y$ equations, and then relating $\mu$ to the
coefficients.  Just as previous work had derived expressions for
$\sum_i\mu_i$, Witt \& Mao found expressions for $\sum_i\mu_ix_i^k$
and $\sum_i\mu_iy_i^k$.  They called these the ``$k^{\rm th}$
moments'', and we adopt their terminology here.  For example, they
found that the first moment $\sum_i\mu_i{\vec\Theta}_i=2{\vec\beta}$
for the SIE model.

The general pattern seen in previous work is that both the total 
magnification and the higher moments can be expressed in terms 
of the model parameters, and that progressively higher moments have
progressively more complicated forms involving more of the parameters.
The expressions' independence of certain
parameters suggests some sort of invariant, but clearly not a 
topological invariant since certain models seemed not to obey 
any magnification relations whatsoever.  Indeed, the origin of the
magnification relations and their absence in certain models has been
a mystery.  In this paper, we provide an explanation of these 
relations, and additionally find a method easier than elimination
theory to derive them.

\subsection{Residue integrals}

As noted above, the lens equations have multiple discrete solutions, 
both real and complex.  While only the real solutions have physical
meaning, it is instructive to consider the complex solutions as well.
In this paper, we will henceforth treat the image coordinates 
${\vec\Theta}$ as complex variables.
We are interested in the sum over these discrete points of various 
quantities, such as the signed magnification, or magnification times
position, etc.  From complex analysis, we know that one may
relate a sum over discrete points to an integral over a contour encircling
those points, by choosing an integrand which has poles at those points.  
For lensed images, which are stationary points of the time delay, there 
is an obvious class of integrands, namely rational functions of the form 
\begin{equation}
f(x,y) = {{g(x,y)}\over{\partial_x\tau\partial_y\tau}}.
\end{equation}
There are complications, which we discuss later, due to the fact that
the integrals here are multidimensional; however, the analogy to the
one dimensional case should be clear.  We need only find the
appropriate function $g$ such that $f$ will have a residue equaling
the quantity we wish to sum over the images, and choose a contour
large enough to enclose all the images.  Now, converting a discrete
sum to a contour integral wouldn't seem to be much progress, however
we can use another idea from one dimensional complex analysis.  Recall
that by inverting coordinates (mapping the origin to infinity and vice
versa) one can see that the sum of the residues of poles inside the
contour is equal to the sum of the residues of poles outside the
contour, but with opposite sign.  In our case, we are summing over all
the finite solutions, so the only pole outside the contour is at
infinity.  This is the essence of the method described in this paper:
we relate the sum over the images to the behavior of the time delay at
infinity, and we simply evaluate the residue at the point(s) at
infinity.  The validity of the resulting magnification relations does
not depend on the image coordinates being real, although of course
their physical applicability does.

In the following section, we further utilize ordinary complex analysis
of one variable and derive trace formulas for those models for which
it is easy to eliminate all but one variable.  In Section
\ref{sec:residue}, we consider the general case, and describe the
basics of multivariable residue calculus.  In Section
\ref{sec:example} we illustrate with specific examples.

%File lens3, section 3 of paper.

\section{Trace Methods}
\label{sec:trace}

In this paper, we shall focus upon models with polynomial lens equations.
Algebraic functions, such as $n$th roots, can be accommodated by 
introducing an additional variable for each algebraic function along
with the polynomial equation it satisfies.
For example, an equation containing $\sqrt{x^3+1}$ is handled by 
introducing $z$ satisfying $z^2=x^3+1$.
The method of this section is useful when all but one variable can be
conveniently eliminated from the lens equations, e.g. Eq. (\ref{extshear}).
Of course, this is always possible in principle, but it may require
computer implementation of Gr\"{o}bner basis algorithms in practice.
Thus, we assume that the $x$-coordinates of the images are the roots
$x_i$ of a polynomial equation of degree $n$,
\begin{equation}
f(x) = \sum_{i=0}^n a_i x^i = 0.
\end{equation}
We assume that the signed magnification $\mu(x)$ of an image at $x$ is  
given by a rational function,
\begin{equation}  \label{defmu}
\mu = \frac{p(x)}{q(x)},
\end{equation}
where the denominator has no common roots with $f(x)$.
This will necessarily be the case 
if the coordinates are chosen generically, since there will then be at 
most one image at a given $x$-coordinate, whose magnification must be a
single-valued algebraic function of $x$. 
We wish to calculate the total magnification
\begin{equation}
M = \sum_i \mu(x_i) \equiv \Tr \mu,
\end{equation}
where the ``trace" notation will be explained below.
Generically, the roots $x_i$ of $f(x)$ will be distinct; since $M$ is
determined by continuity when some roots coincide we will always
consider the generic case.

Let $A = {\mathbb{C}}[x]$ be the ring of polynomials in $x$ with complex
coefficients.
We call two polynomials $g$ and $h$ equivalent, writing $g \sim h$,  
if they differ by a polynomial multiple of $f(x)$.
This sorts the polynomials into equivalence classes, 
and we denote the class containing $g$ by $[g]$ and the set of all 
equivalence classes by $A_f$.
This is of use for our problem because all polynomials in a given class
take the same values at the $x_i$ and therefore have the same trace.
Addition and multiplication of classes are well-defined by 
$[g]+[h]=[g+h],\; [g][h]=[gh],$ and $A_f$ is itself a ring.

Now we observe that each class $[g] \in A_f$ contains a representative
which has degree (at most) $n-1$.
Indeed, the relation $f \sim 0$ implies
\begin{equation}    \label{reduce}
a_n x^n \sim -\sum_{i=0}^{n-1} a_i x^i,
\end{equation}
and this can be used to eliminate all terms of degree $n$ or 
greater from a polynomial $g$.
The resulting polynomial is unique, being determined by its values at
the $n$ roots $x_i$.
It follows that $A_f$ is in fact a vector space of dimension $n$ over 
$\mathbb{C}$.

Now consider, instead of polynomials, the set $R$ of rational functions of $x$
which are defined at the zeros of $f(x)$, and call two elements equivalent
if their difference is $f(x)$ times another element.
Then all elements of a class have the same trace, and the set $R_f$ of
equivalence classes is again a ring.
In fact, it is isomorphic to $A_f$.
An isomorphism is obtained by associating to a class $[g]$ in $A_f$
the obvious class $[g]$ in $R_f$.
To see that this mapping is invertible, we must find a polynomial
representative $g$ of an arbitrary class $[h]$ in $R_f$.
To do so, simply let $g(x)$ be the unique polynomial of degree $n-1$
satisfying $g(x_i) = h(x_i)$ for all $i$.
Then $g-h$ is a rational function vanishing at every $x_i$, so
when expressed in lowest terms its numerator must be a multiple of $f(x)$.
Therefore $g \in [h]$.

At this point we can explain the ``trace" terminology for the sum over
the roots $x_i$.\cite{CLO98}
The vector space $A_f$ has a basis consisting of (the classes of) the
$n$ polynomials $\delta_i$ of degree $n-1$
defined by $\delta_i(x_j)=\delta_{ij}$.
Fix an element $g \in A_f$ and consider the linear operator on $A_f$
given by multiplication by $g$.
In the given basis, the matrix of this operator is diagonal, with entries
$g(x_i)$.
Hence the trace of this matrix, which is independent of the basis and 
coincides with the trace of the operator, is simply
$\Tr g = \sum_i g(x_i)$.
For example, consider $\Tr x$, which just gives the sum of the roots.
Choosing the basis $\{1,x,x^2,\ldots,x^{n-1}\}$ for $A_f$, the matrix of
the operator of multiplication by $x$ has only one nonzero diagonal entry,
$-a_{n-1}/a_n$, arising from the relation 
$x \cdot x^{n-1} \sim -(a_{n-1}/a_n) x^{n-1} + \cdots$.
This recovers the standard result for the sum of the roots of a polynomial
and shows how our method generalizes others based on that result
\cite{wm95,rhie,dalal,wm99}.

Returning to the problem of computing the total magnification, we see that
$M = \Tr \mu = \Tr [\mu]$ can be computed using any element in its equivalence
class.
For example, we can choose the unique polynomial representative of 
degree $n-1$.
Of course, we do not determine this polynomial from its values $\mu(x_i)$,
since we do not know the $x_i$ explicitly.
Instead, we seek a polynomial solution of degree $n-1$ to the condition
(\ref{defmu}) defining $[\mu]$,
\begin{equation}
[\mu q(x)] = [p(x)].
\end{equation}
This is solved by using Eq. (\ref{reduce}) to reduce the degree of each
side to $n-1$, and then equating coefficients.

To compute the trace we use a formula due to Euler, which we derive by
means of the residue theorem for complex contour integrals.
A purely algebraic proof is not difficult \cite{W63}, 
but our derivation shows the relevance of residue methods and
motivates the multivariable generalization which
we describe in the following section.
Consider the contour integral,
\begin{equation}
\frac{1}{2\pi i} \oint \frac{f'(x) \mu}{f(x)}\, dx,
\end{equation}
where the contour is a large circle in the complex plane enclosing all the
zeros $x_i$ of $f(x)$.
The integrand has a pole of residue $\mu(x_i)$ at $x_i$, and
consequently the integral is $\Tr \mu$.
However, we can also regard the contour as encircling the point at infinity
and evaluate the integral in terms of the residue there, introducing if we
wish the new variable $u=1/x$ to move the point at infinity to the origin.
Furthermore, the integral is unchanged if $f'(x)\mu$ is replaced by any
member of its equivalence class.
By choosing the polynomial representative of degree $n-1$ we need only
evaluate
\begin{equation}
\frac{1}{2\pi i} \oint \frac{x^k}{f(x)} dx = \frac{\delta_{k,n-1}}{a_n}, 
\;\;\; 0 \leq k \leq n-1.
\end{equation}
This proves Euler's formula,
\begin{equation}
\Tr \mu = \frac{\mbox{coefficient in degree}\; n-1 \;\mbox{of}\; f'(x)\mu}
        {\mbox{coefficient in degree}\; n \;\mbox{of}\; f(x)},
\end{equation}
where it is understood that the polynomial representative of degree
$n-1$ is meant in the numerator.
This makes it clear that the total magnification is determined by the
leading behavior of $f(x)$ and $f'(x)\mu$ at infinity.

As an example, we consider a generalization of the SIS + elliptical 
potential ($n=1$ multipole) with an arbitrary harmonic, 
$\psi = br + \gamma br \cos m\theta$.
As an aid to clarity, we shall depart from conventional notation,
and instead rewrite the lens equations so that
the variables are $x,y$, and parameters are denoted $a,b,c,\ldots$.
In this case we define 
$x=e^{i\theta}, y=r, a=\gamma, b=b, c=s, d=e^{i\theta_s}$.
The lens equations then take the form,

\begin{eqnarray}
f(x) = mabx^{2m} + cd^{-1} x^{m+1} - cd x^{m-1} - mab & = & 0, \\ 
2(y-b) - c(xd^{-1} + dx^{-1}) - ab(x^m + x^{-m}) & = & 0,
\end{eqnarray}
with the magnification satisfying
\begin{equation}
[m^2 ab (x^m + x^{-m}) + c(xd^{-1} + dx^{-1})] \mu = 2y.
\end{equation}
Eliminating $y$ results in

\begin{eqnarray}
[m^2 ab x^{2m} & + & cd^{-1} x^{m+1} + cd x^{m-1} + m^2 ab] \mu  = \\ \nonumber 
ab x^{2m} & + & cd^{-1} x^{m+1} + 2b x^m + cd x^{m-1} + ab.   
\end{eqnarray}
Dividing through by $x$ and replacing $x^{-1}$ in the resulting
equation with a polynomial equivalent via the relation $x^{-1}f(x) \sim 0$
produces

\begin{eqnarray}
[2m^2 ab x^{2m-1} & + & (m+1)cd^{-1} x^m - (m-1)cd x^{m-2}] \mu = \\ \nonumber
2ab x^{2m-1} & + & (1+m^{-1})cd^{-1} x^m + 2b x^{m-1} + (1-m^{-1})cd x^{m-2},
\end{eqnarray}
where the left side is precisely in the Euler form $f'(x)\mu$.
Then we immediately have the total magnification invariant as
\begin{equation}
M = \Tr \mu = \frac{2ab}{mab} = \frac{2}{m}.
\end{equation}
It is no harder to verify that the total magnification is the same for
a potential containing an arbitrary finite sum of harmonics 
$\sum_{k=1}^m b_k r \cos k\theta$;
the highest harmonic determines $M$.  Other models are amenable to this
method as well.

In principle one can compute moments by this method, by replacing 
$\mu$ with $x^k \mu$ in the relevant equations, but we have preferred
the residue methods of the next section for moment computations.

%File lens4, section 4 of paper.

\section{Residue Methods: Theory}
\label{sec:residue}

Let us review the key steps in our contour integral derivation of Euler's
formula in the previous section.
First, we expressed the total magnification as a complex contour integral
of a function having poles at the image locations.
Second, we converted this to an integral around the point at infinity.
This amounts to viewing the complex plane as a subset of the Riemann sphere,
or complex projective space \cp{1}.
Our change of variables $u=1/x$ connects two coordinate charts on \cp{1}
centered at the origin and at infinity.
Finally, we evaluated the residue at infinity, making it clear that the
total magnification depends on the behavior of the integrand at infinity.

These steps all have multivariable analogs.
In fact, there is a well-developed, if little-known, residue theory for
meromorphic differential forms in several complex variables.
This makes it possible to compute directly the total magnification and 
moments for lens models without reducing to one-variable lens equations.
The theory is particularly effective in the situation of two variables, and
we describe it in this case.
References include [\onlinecite{ATY94,AY83,BK86,GH78}].
An application to chemical reaction rate equations appears in 
Ref. [\onlinecite{ABKY83}],
but we are not aware of other physical applications in the literature.

We consider a meromorphic two-form 
\begin{equation}
\omega = \frac{g(x,y)\, dx\, dy}{P_1(x,y) P_2(x,y)},
\end{equation}
on ${\mathbb{C}}^2$, which we view as a subset of the compact complex projective
space \cp{2}.
Here $P_1,P_2$ are polynomials having finitely many common zeros (the
image locations) of multiplicity one (as is generically the case), 
and $g$ is also a polynomial.
Such a form can be integrated over a 2-cycle, a compact two-dimensional
real submanifold of \cp2.
Since $\omega$ is closed ($d\omega=0$), the integral depends only on
the homology class of the cycle.
For example, in a small neighborhood of a common zero of the $P_i$, we can
integrate over the ``torus" $T: \{|P_1| = |P_2| = \epsilon \}$, defining the
residue of $\omega$ at this zero:
\begin{equation}
{\rm Res~} \omega = \left( \frac{1}{2\pi i} \right)^2 \int_T \omega.
\end{equation}
[The standard orientation of $T$ is that specified by the nonvanishing 2-form
$d(\arg P_1)\, d(\arg P_2)$.
That is, $T$ is oriented so that $dP_1\,dP_2/(2\pi i)^2 P_1 P_2$ has
positive integral.]
We will always denote by $J$ the naive Jacobian of the mapping 
$(x,y) \mapsto [P_1(x,y),P_2(x,y)]$, namely
\begin{equation}
J = \frac{\partial (P_1,P_2)}{\partial(x,y)} = 
\left| \begin{array}{cc} \partial_x P_1 & \partial_y P_1 \\ 
\partial_x P_2 & \partial_y P_2 \\ \end{array} \right|.
\end{equation}
This coincides with the physical Jacobian relating corresponding area elements
in the source and image planes if $x,y$ are rectangular coordinates,
and $P_1,P_2$ are the corresponding lens equations,
but requires a correction factor otherwise.
The first key fact we need is that the residue at a
nondegenerate zero (one where $J$ does not vanish), located
say at the origin, is given by\cite{GH78}
\begin{equation}     \label{simple}
{\rm Res~} \frac{g\, dx\, dy}{P_1 P_2} = \frac{g(0)}{J(0)},
\end{equation}
which is equal to the magnification of the corresponding image if $g$ is
chosen appropriately ($g=1$ for rectangular coordinates).
Moments of magnification can be computed by including additional
monomial factors in $g(x,y)$.

Next we need the Global Residue Theorem \cite{GH78}, which states that
the sum of all the residues of a meromorphic form, such as $\omega$, on any
compact manifold, such as \cp{2}, vanishes.
Note that this sum is over the common zeros of the $P_i$ only,
so that points where only one polynomial vanishes do not contribute;
also the points summed over may depend on the choice of the factorization
$P_1P_2$ of the denominator of $\omega$.
The theorem makes it possible to replace the sum over the residues at 
the common zeros in ${\mathbb{C}}^2$ by minus the sum of residues at points 
at infinity in \cp{2}.
This is the fundamental explanation for the existence of magnification
relations in general: compactness relates the sum of finite residues
to the behavior of the lens equations at infinity, indeed to 
a finite number of terms in an expansion around infinity. 
It remains to explain how to locate the common zeros at infinity and compute
their residues.

\cp{2} is conveniently described by homogeneous coordinates 
$[X,Y,U] \neq [0,0,0]$,
where $[\lambda X, \lambda Y, \lambda U]$ is identified with $[X,Y,U]$
for all complex $\lambda \neq 0$.
The points with $U \neq 0$ can be represented in the form $[x,y,1]$ and
are viewed as the subset ${\mathbb{C}}^2$ of finite points.
The polynomials $P_i(x,y)$ correspond to homogeneous polynomials 
$P_i^h(X,Y,U) \equiv U^{\deg P_i} P_i(X/U,Y/U)$.
Their common zeros at infinity are those having $U=0$.
These also lie in coordinate charts diffeomorphic to ${\mathbb{C}}^2$, 
described by $[1,y,u]$ or $[x,1,u]$.
In these charts they can be treated just as ``finite" zeros are.
The meromorphic form $\omega$ is homogenized so as to have total degree zero,
that is, 

\begin{eqnarray}
\omega^h & = & 
\frac{g(X/U,Y/U) \, d(X/U) \, d(Y/U)}{P_1(X/U,Y/U) P_2(X/U,Y/U)} \\ \nonumber
& = & \frac{g^h U^{\deg P_1 + \deg P_2 - \deg g - 3} 
(U dX dY - X dU dY - Y dX dU)}{P_1^h P_2^h}.
\end{eqnarray}
When $n = \deg g - \deg P_1 - \deg P_2 + 3 > 0$, the denominator
of $\omega^h$ takes the form $U^n P_1^h P_2^h$.
This creates a subtlety in that the above theory applies to
a chosen factorization of the denominator into two factors.
Thus, we may factor it as $(U^n P_1^h) (P_2^h)$ and treat these as the two
factors in applying the residue theorem.
The common zeros then consist of all (finite and infinite) common zeros of
the $P_i^h$, together with any zeros of $P_2^h$ alone at infinity.
The latter might have been overlooked in a naive application of the
theorem.
In the examples we will consider in detail, no such additional zeros exist, 
but we will point out a case where they do.

Unfortunately, the zeros at infinity are rarely nondegenerate, 
and their residues cannot be computed using Eq. (\ref{simple}).
Instead, they typically lie at singular points of the curves $P_i=0$,
that is, at least one curve has 
multiple branches meeting at this point.\footnote{Note that these
curves have one complex, but two real dimensions, and so are real surfaces.}
There is a classical method, dating back to Newton, for finding the branches
of an algebraic curve $P(x,y)=0$ at a singular point, taken to be the origin.
The branches are given as Puiseux series, or fractional power series,
of the form 
\begin{equation}
y = \sum_{i=0}^\infty a_i x^{\alpha_i},
\end{equation}
where the exponents $\alpha_i$ form an increasing sequence of rational
numbers whose denominators eventually stabilize.
The possibilities for the leading exponent $\alpha_0$ are determined 
by requiring 
that at least two terms in the polynomial $P(x,a_0 x^{\alpha_0})$ have 
the same degree, while the remaining terms have higher degree.
Then $a_0$ is found by demanding the vanishing of the terms of minimal
degree.
An elegant graphical method for identifying the possible exponents $\alpha_0$
is provided by the Newton polygon, or diagram \cite{W50,BK86}.
For each monomial $x^a y^b$ appearing in $P(x,y)$, plot the point $(a,b)$ 
in the coordinate plane.
Begin at the lowest of the leftmost points (minimize $a$, then minimize $b$)
and draw a polygonal path with vertices at a subset of the points, 
terminating at the leftmost of the lowest points (minimize $b$,
then minimize $a$), and choosing each successive segment to have the
smallest possible slope (steepest possible negative slope).
This is the Newton diagram of the polynomial $P(x,y)$.
The points lying on any segment of the Newton diagram represent terms in
$P$ which will have minimal degree if $\alpha_0$ is chosen as the negative
reciprocal of the slope of that segment.
The next term in the Puiseux series can be found by applying the same
procedure to $P(x,a_0x^{\alpha_0} + \tilde{y})$, and so on.
One term is often enough for computing the total magnification; higher
moments require more terms in general.
The denominators of the exponents stabilize at the stage where the Newton
diagram has only a single segment.
The curve $P(x,y)=0$ has at least as many branches at the origin as there
are segments in the Newton diagram, and has more if there are multiple 
solutions to the equations for the coefficients $a_i$.

Consider one particular branch $X$ of the curve $P_1(x,y)=0$ at a singular
point, given by a Puiseux series $y=a_0x^{m/n} + \cdots$, with $m,n$
relatively prime.
On this branch draw a small circle $C$ around the origin; its projection
on the $x$-plane must wind $n$ times around the origin. 
Now construct a 2-torus $\delta C$ by ``thickening" $C$: 
at any point $p$ of $C$ take a 
plane transverse to $X$ and a small circle in this plane with center $p$.
As $p$ moves around $C$ this circle sweeps out the torus $\delta C$.
We may construct such a torus for each branch of the curve.
These are called Leray tori, and the ``thickening" operator $\delta$
is the Leray coboundary.

Our objective is to compute the residue integral $\int_T \omega$ at
the origin.
We can work entirely within a small ball $B$ around the origin.
The residue depends on the homology class of $T$ in 
$H_2(B - \{P_1P_2=0\})$.
As we explain in the Appendix, this class is, up to sign, the sum of the classes
of the Leray tori constructed on the branches of {\it either} of the curves
$P_1=0$ or $P_2=0$.
Therefore the residue is the sum of the integrals of $\omega$ over either
set of Leray tori, with appropriate orientation.

The integral over a Leray torus $\delta C$ lying on a branch of $P_1(x,y)=0$
given by a Puiseux series $y=p(x)$ is computed using the Leray residue 
formula.
We give this formula under the assumption that $\partial_y P_1$ does
not vanish on the given branch, which amounts to assuming that 
$y=p(x)$ is a branch of an irreducible factor of $P_1$ which appears
to the first power only (the analog of a simple rather than multiple pole).
The general formula can be found in Ref. [\onlinecite{AY83}].
Our case reads
\begin{equation}
\frac{1}{2\pi i} \int_{\delta C} \frac{g\;dx\;dy}{P_1 P_2} = 
- \int_C \frac{g\;dx}{P_2 \, \partial_y P_1} \Big|_{y=p(x)}
\end{equation}
which is proved as follows.
To integrate over $\delta C$, we can integrate first over the circles in
the planes transverse to $C$, then over $C$ itself.
Since these circles are centered at $P_1=0$, 
we can change variables from $x,y$ to $x,P_1$ by means of
$dP_1\;dx = - dx\;dy\;\partial_y P_1$ and then integrate over $P_1$
by means of the one-variable residue theorem.
The evaluation of the residue at $P_1=0$ by substituting $y=p(x)$ 
leaves another one-variable residue integral to be performed.
For this one must keep in mind that the cycle $C$ may wind around the
origin several times in the $x$-plane.

The Leray residue formula holds with obvious notational and possible sign
changes if the Leray torus lies on a branch of $P_2(x,y)=0$, or if we 
choose to eliminate $x$ rather than $y$ in favor of a $P_i$.

%File lens5, section 5 of paper

\section{Residue Methods: Examples}
\label{sec:example}

In this section we apply the residue methods just developed  
to various lens models.
The first is a generalized $n=2$ multipole model with potential
$\psi = br + (\gamma /2)r^2 \cos m\theta$.
We again redefine notation to clarify the computation, writing
the variables as $x,y$ and the parameters as $a,b,c,\ldots$.
Let $x=e^{i\theta}, y=r, a=\gamma, b=b, c=s, d=e^{i\theta_s}$.
The lens equations then take the form,
\begin{eqnarray}
P_1 & = & cd^{-1}x^{m+1} - cdx^{m-1} + \frac{m}{2}ay(x^{2m}-1) = 0, \\
P_2 & = & 2(y-b)x^m - cd^{-1}x^{m+1} - cdx^{m-1} -ay(x^{2m}+1) = 0.
\end{eqnarray}
Because $x,y$ are not rectangular coordinates, there is an extra Jacobian
factor in the magnification, and we find
\begin{equation}
\mu = \frac{4yx^{2m-1}}{J}.
\end{equation}
Consequently the total magnification is given by the residue sum, 
\begin{equation}
M = \sum_{\rm images} {\rm Res~} \frac{4yx^{2m-1} dx \, dy}{P_1P_2}.
\end{equation}
The residue theorem relates this to the sum of the residues at points at
infinity in \cp2, which are found from the homogeneous forms of the $P_i$:
\begin{eqnarray}
P_1^h & = & cd^{-1}X^{m+1}U^m - cdX^{m-1}U^{m+2} + 
            \frac{m}{2} aY(X^{2m}-U^{2m}), \\
P_2^h & = & 2X^m(YU^m-bU^{m+1}) - cd^{-1}X^{m+1}U^m - cdX^{m-1}U^{m+2} 
-aY(X^{2m}+U^{2m}).
\end{eqnarray}

The common roots at infinity are those with $U=0$, and there are two:
$[X,Y,U] = [1,0,0],$ and $[0,1,0]$.
The total magnification is minus the sum of the residues at these points of
\begin{equation} \label{shearres}
\frac{ 4YX^{2m-1}U^{2m+2}\,d(X/U)\,d(Y/U) }{P_1^h P_2^h}.
\end{equation}

Consider first the point $[1,0,0]$, which we examine in the affine chart
$[X,Y,U] = [1,y,u]$, where

\begin{eqnarray}
P_1^{X=1} & = & cd^{-1}u^m - cdu^{m+2} + \frac{m}{2}ay(1-u^{2m}), \\
P_2^{X=1} & = & 2u^m(y-bu) - cd^{-1}u^m - cdu^{m+2} - ay(1+u^{2m}),
\end{eqnarray}
and we need the residue at the origin of 
$4yu^{2m-1}\,du\,dy/P_1^{X=1}P_2^{X=1}$.

The Newton diagram for $P_1$ is shown in Figure 1; there is a single branch
on which to leading order $y \sim u^m$, as is easily verified by solving
$P_1^{X=1} = 0$ for $y$.
The Leray formula evaluates the residue as the one-variable residue of
\begin{equation}  \label{shearresx1}
- \frac{ 4yu^{2m-1} du }{ P_2^{X=1} \partial_y P_1^{X=1} },
\end{equation}
where $y \sim u^m$.
But it is easily seen that the leading behavior of this 1-form near $u=0$ is
$u^{2m-1}\,du$, so that there is no pole and no
residue for $m>0$.

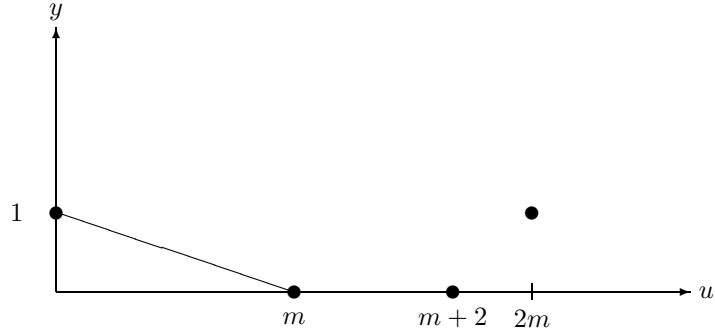
\begin{figure}
\begin{picture}(240,100)(0,-10)
\put(0,0){\vector(0,1){100}}
\put(0,0){\vector(1,0){240}}
\put(0,30){\circle*{5}}
\put(90,0){\circle*{5}}
\put(150,0){\circle*{5}}
\put(180,-3){\line(0,1){6}}
\put(180,30){\circle*{5}}
\put(0,30){\line(3,-1){90}}
\put(0,106){\makebox(0,0){$y$}}
\put(246,0){\makebox(0,0){$u$}}
\put(-15,30){\makebox(0,0){$1$}}
\put(90,-10){\makebox(0,0){$m$}}
\put(150,-10){\makebox(0,0){$m+2$}}
\put(180,-10){\makebox(0,0){$2m$}}
\end{picture}
\caption{The Newton diagram for $P_1^{X=1}$ in the $n=2$ multipole model.
The degree of a monomial in $y$ ($u$) is plotted vertically (horizontally).
The case $m=3$ is shown.}
\end{figure}

We examine the remaining root $[0,1,0]$ in the chart
$[X,Y,U] = [x,1,u]$ where

\begin{eqnarray}
P_1^{Y=1} & = & cd^{-1}x^{m+1}u^m - cdx^{m-1}u^{m+2} + 
                \frac{m}{2}a(x^{2m}-u^{2m}), \\
P_2^{Y=1} & = & 2x^mu^m(1-bu) - cd^{-1}x^{m+1}u^m - cdx^{m-1}u^{m+2} 
- a(x^{2m}+u^{2m}),
\end{eqnarray}
and we compute 
\begin{equation}
{\rm Res~} \frac{ 4x^{2m-1}u^{2m-1}\,dx\,du }{P_1^{Y=1}P_2^{Y=1}}.
\end{equation}

The Newton diagram for $P_1^{Y=1}$ is shown in Figure 2, and gives the leading
behavior $u = lx + \cdots$.
With this behavior, the lowest-order terms of $P_1^{Y=1}$ vanish iff 
$l^{2m}=1$, so that there are $2m$ branches with 
$l_p = \exp(ip\pi/m),\;p=1,\ldots,2m$.

With the Leray formula, the total magnification becomes
\begin{equation}   \label{shearresy1}
\sum_{u=l_px+\cdots} 
- {\rm Res~} \frac{ 4x^{2m-1}u^{2m-1}\,dx }{ P_2^{Y=1} \partial_u P_1^{Y=1} }.
\end{equation}
The leading terms are readily identified, and indeed the behavior is as
$dx/x$, with residue
\begin{equation}
\sum_{p=1}^{2m} \frac{2}{m^2 a(l_p^m-a)} = 
\frac{2}{m^2 a} \sum_{p=1}^{2m} \frac{1}{(-1)^p-a} = \frac{4}{m(1-a^2)}.
\end{equation}
In terms of the original parameters of the model, 
\begin{equation}
M = \frac{4}{m(1 - \gamma^2)},
\end{equation}
an attractive generalization of the known result\cite{dalal} for $m=2$.

\setlength{\unitlength}{0.75pt}
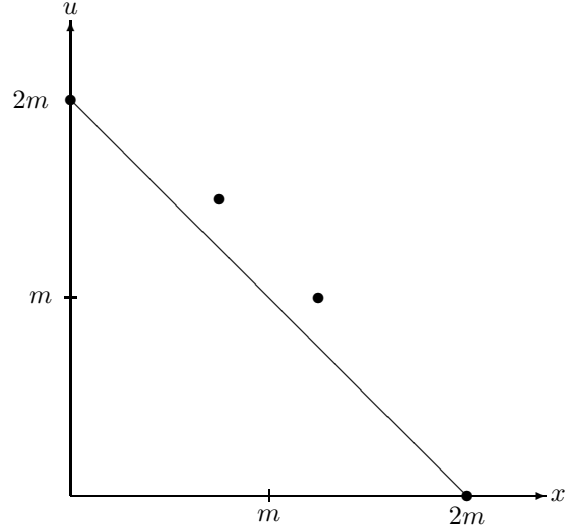
\begin{figure}
\begin{picture}(240,240)(0,-10)
\put(0,0){\vector(0,1){240}}
\put(0,0){\vector(1,0){240}}
\put(0,200){\circle*{5}}
\put(-3,100){\line(1,0){6}}
\put(200,0){\circle*{5}}
\put(100,-3){\line(0,1){6}}
\put(75,150){\circle*{5}}
\put(125,100){\circle*{5}}
\put(0,200){\line(1,-1){200}}
\put(0,246){\makebox(0,0){$u$}}
\put(246,0){\makebox(0,0){$x$}}
\put(-15,100){\makebox(0,0){$m$}}
\put(-20,200){\makebox(0,0){$2m$}}
\put(100,-10){\makebox(0,0){$m$}}
\put(200,-10){\makebox(0,0){$2m$}}
\end{picture}
\caption{The Newton diagram for $P_1^{Y=1}$ in the $n=2$ multipole model.}
\end{figure}

Once the branches of the $P_i$ have been identified, it is easy to modify
the calculation to compute moments rather than total magnification.
For example, let us compute the first $x$-moment, $\Tr x\mu$.
(We continue to use the $\Tr$ notation for the sum over the images.)
Since $x=e^{i\theta}$ in terms of the physical variables of this model,
from the real and imaginary parts of the result we can obtain the moments
weighted by $\cos \theta$ and $\sin \theta$, in the case that all images
are real.
In homogeneous coordinates this simply gives an additional factor 
$X/U$ in the residue in Eq. (\ref{shearres}).
At the point $[1,0,0]$ this adds a factor $1/u$ to the one-variable residue
(\ref{shearresx1}), changing the behavior to $u^{2m-2}\,du$.
There is still no contribution for $m \ge 1$.
At the point $[0,1,0]$, there is an extra factor $x/u = l^{-1} + \cdots$
in the residue (\ref{shearresy1}), changing the contribution to
\begin{equation}
\sum_p \frac{2}{m^2al_p(l_p^m-a)} = \frac{2}{m^2a} \sum_{p=1}^{2m}
\frac{e^{-ip\pi /m}}{(-1)^p-a},
\end{equation}
which can be evaluated in closed form if desired.

From the real part of $\Tr yx\mu$ we can obtain the moment of $r \cos \theta$,
the physical $x$-coordinate of the image.
This leads to a factor $YX/U^2$, which worsens the singularity at $[0,1,0]$
to a double pole, requiring an additional term in the Puiseux expansion to
obtain the residue.
The result, for the true external shear model $m=2$, is
\begin{equation}
\Tr \mu r \cos \theta = -\frac{2 x_s}{(1-\gamma)(1-\gamma^2)}.
\end{equation}

This example illustrates the general situation.
Higher moments produce extra monomial factors in the residue expression.
In general this will worsen the singular behavior at points at infinity,
although this may not occur for certain branches, such as those at $[0,1,0]$
in the example of $\Tr x\mu$.
This will have two effects: points which do not contribute to the total 
magnification generally will contribute to higher moments, and more terms
in the Puiseux expansions will be required for higher moments.
Both effects will result in higher moments being given by more complex
expressions, with more model parameters contributing.

We turn to a second example, the Singular Isothermal Ellipse (SIE)
potential\cite{bk87,wm99}.
For this model $\psi = bR = b\sqrt{x^2 + y^2q^{-2}}$, where $x,y$ are
rectangular coordinates and $b,q$ are parameters.
The lens equations are

\begin{eqnarray}
\tau_x & = & x - x_s - \frac{bx}{R} = 0, \\
\tau_y & = & y - y_s - \frac{by}{q^2R} = 0 ,
\end{eqnarray}
and the magnification is given by
\begin{equation}
\mu^{-1} = \left| \begin{array}{cc} 
\tau_{xx} & \tau_{xy} \\ \tau_{xy} & \tau_{yy} 
\end{array} \right|.
\end{equation} 
Algebraic manipulation leads to the polynomial equations,
\begin{eqnarray}
p_1 & = & q^2x\tau_y - y\tau_x = q^2x(y-y_s) - y(x-x_s) = 0, \\
p_2 & = & R^2\tau_x^2 - 2bRx\tau_x = (x-x_s)^2 (x^2+y^2q^{-2}) - b^2x^2 = 0.
\end{eqnarray}
However, these equations have the extraneous solution $(x,y) = (0,0)$,
which does not satisfy the original lens equations.
This can be eliminated by substituting $y=wx$, and adopting the
modified equations
\begin{eqnarray}
P_1 & = & p_1/q^2x = wx - y_s - \eta wx + \eta wx_s, \\
P_2 & = & p_2/x^2 = b^2 - (x-x_s)^2 (1+ \eta w^2),
\end{eqnarray}
where we have set $\eta = q^{-2}$.
Relating the naive Jacobian $J$ of the $P_i$ with respect to $x,w$ to the
Hessian of $\tau$ gives the magnification in the new variables,
\begin{equation}
\mu = \frac{2x(x-x_s)(1 + \eta w^2)}{J}.
\end{equation}

The moments of the magnification with respect to $x$ are given by
\begin{equation}
\Tr x^k \mu = \sum {\rm Res~} 
\frac{2x^{k+1}(x-x_s)(1 + \eta w^2)\,dx\,dw}{P_1P_2}.
\end{equation}
In homogeneous coordinates we have
\begin{eqnarray}
P_1^h & = & (1-\eta) XW - y_s U^2 + \eta x_s WU,   \\ 
P_2^h & = & b^2 U^4 - (X- x_sU)^2 (U^2 + \eta W^2),
\end{eqnarray}
and we need to compute
\begin{equation}
\Tr \mu x^k = \sum {\rm Res~} \frac{ 2X^{k+1}U^{2-k} (X-x_sU) (U^2 + \eta W^2) 
\,d(X/U)\,d(W/U) } {P_1^h P_2^h}.
\end{equation}
There are two common zeros of the $P_i^h$ at infinity, namely 
$[X,W,U] = [1,0,0],\,[0,1,0]$, and in the corresponding affine charts 
we have
\begin{eqnarray}
P_1^{X=1} & = & (1-\eta) w - y_s u^2 + \eta x_s wu,  \\
P_2^{X=1} & = & b^2 u^4 - (1 - x_s u)^2 (u^2 + \eta w^2),  \\
P_1^{W=1} & = & (1-\eta) x - y_s u^2 + \eta x_s u,  \\
P_2^{W=1} & = & b^2 u^4 - (x - x_s u)^2 (u^2 + \eta).
\end{eqnarray}
The residues at these zeros may be computed via the Leray formula applied
to the branches of either $P_1$ or $P_2$; we have done both computations
and the latter seems slightly simpler. 
In each case the branches can be determined directly without appealing to
the Newton diagrams.
At $[0,1,0]$, the equation $P_2^{W=1} = 0$ is solved by
\begin{equation}
x = x_s u \pm b \eta^{-1/2} u^2 (1 + \frac{u^2}{\eta})^{-1/2} ,
\end{equation}
giving one branch for each choice of sign.
The contribution to $\Tr \mu x^k$ from one branch is
\begin{equation}
{\rm Res~} \frac{2x^{k+1}(x - x_s u)(u^2 + \eta) \,dx\,du}{u^{k+1}
P_1^{W=1} P_2^{W=1}}.
\end{equation}
Applying the Leray formula gives
\begin{equation}
{\rm Res~} \frac{2x^{k+1}(x - x_s u)(u^2 + \eta) \,du}{u^{k+1}
P_1^{W=1} \partial_x P_2^{W=1}},
\end{equation}
which simplifies to 
\begin{equation}
{\rm Res~} \frac{x^{k+1} \,du}{u^{k+1} [x - y_s u^2 - \eta (x - x_s u)]} =
x_s^k,
\end{equation}
where only the first term in the series expansion of $x$ was required.
The two branches at this point thus contribute $2x_s^k$ to the moment
of $x^k$.

At the remaining point $[1,0,0]$, $P_2^{X=1} = 0$ is solved by
\begin{equation}
w = \pm \frac{iu}{\sqrt{\eta}} \sqrt{ 1 - \frac{b^2u^2}{(1 - x_s u)^2}} =
\pm \frac{iu}{\sqrt{\eta}} ( 1 - \frac{1}{2} b^2u^2 + \cdots ),
\end{equation}
and we need
\begin{equation}
{\rm Res~} \frac{2(1 - x_s u)(u^2 + \eta w^2) \, du \, dw} {u^{k+1}
P_1^{X=1} P_2^{X=1} },
\end{equation}
summed over the two branches.
Integrating over $w$ using the Leray residue formula and simplifying yields
\begin{equation}
{\rm Res~} \frac{(u^2 + \eta w^2) \, du}{\eta w u^{k+1} (1 - x_s u) 
[(1-\eta)w + \eta x_s wu - y_s u^2]},
\end{equation}
where $u^2 + \eta w^2 = b^2 u^4 + \cdots$.
Since $w \sim u$, there is no pole for $k< 2$: the total magnification and
first moment are simply given by the contributions from $[0,1,0]$.
There is an additional contribution to the second moment given by
$-2b^2/(1-\eta^2) = 2b^2q^2/(1-q^2)$, which agrees with the result of 
Ref. [\onlinecite{wm99}].

We have verified the results of \textcite{wm99} through the third moment.
Those authors noted that the moments of $x$ were independent of $y_s$ to
this order, and that ``this seems remarkable" in view of the dependence
of $\mu$ on this parameter.
The explanation is that the term $y_s u^2$ in $P_1^{X=1}$
does not contribute to the residue for low moments; 
indeed it {\it does} contribute to the residues
for the third moment but its contribution cancels between the two branches.
The fourth and higher moments do depend on $y_s$.

Lastly, we note that it is entirely straightforward, and no more
work, to generalize the calculation to include an 
arbitrarily oriented external shear term.  The first moment, for
example, takes the form 
\begin{equation}
\Tr\mu x = {2\over{(1-\gamma^2)^2}}(x_s + \gamma_1 x_s + \gamma_2 y_s),\qquad
\Tr\mu y = {2\over{(1-\gamma^2)^2}}(y_s + \gamma_2 x_s - \gamma_1 y_s)
\end{equation}
where $\gamma_1\equiv\gamma\cos 2\theta_\gamma$,
$\gamma_2\equiv\gamma\sin 2\theta_\gamma$, and $\theta_\gamma$ is the
orientation angle of the shear (see Table \ref{tab1}).

As our final example we consider microlensing due to a collection of $N$
coplanar point masses.
We adopt Witt's complex notation, writing $z=x+iy$ for the position
of an image and $w$ for the position of the source.
The lenses have masses $m_i$ and positions $z_i$.
The lens equations are 

\begin{equation}
z - w - \sum_{i=1}^N \frac{m_i}{\bar{z} - \bar{z}_i} = 0,
\end{equation}
and its complex conjugate; when we complexify the coordinates $x,y$
of an image, $z$ and $\bar{z}$ become independent variables and the
conjugate equation becomes an independent condition as well.
The observable (real) images are those for which $\bar{z}$ is the conjugate
of $z$.
Clearing denominators, we set

\begin{eqnarray}
P_1 & = & (z-w) \prod_i (\bar{z} - \bar{z}_i) - \sum_i m_i \prod_{j \neq i}
(\bar{z} - \bar{z}_j),  \\ 
P_2 & = & (\bar{z} - \bar{w}) \prod_i (z - z_i) - \sum_i m_i \prod_{j \neq i}
(z - z_j).
\end{eqnarray}
For the magnification we find
\begin{equation}
\mu = J^{-1} \prod_i (z - z_i)(\bar{z} - \bar{z}_i),
\end{equation}
where $J = \partial(P_1,P_2)/\partial(z,\bar{z})$.
For the $k$th moment of magnification, we must compute
\begin{equation}
\Tr \mu z^k = {\rm Res~} \frac{z^k \prod_i (z - z_i)(\bar{z} - \bar{z}_i) 
\, dz \, d\bar{z} } {P_1 P_2}.
\end{equation}
The homogeneous polynomials are
\begin{eqnarray}
P_1^h & = & (Z - wU) \prod_i (\bar{Z} - \bar{z}_i U) - 
\sum_i m_i U^2 \prod_{j \neq i} (\bar{Z} - \bar{z}_j U),  \\ 
P_2^h & = & (\bar{Z} - \bar{w}U) \prod_i (Z - z_i U) -
\sum_i m_i U^2 \prod_{j \neq i} (Z - z_j U).
\end{eqnarray}
Setting $U=0$, we obtain $P_1^h = Z \bar{Z}^N, \, P_2^h = \bar{Z} Z^N$,
so there are two common zeros at infinity, 
$[Z,\bar{Z},U] = [1,0,0], \, [0,1,0]$.
However, we can say more: each of these zeros has multiplicity $N$.
Since the homogeneous polynomials each have degree $N+1$, the number of finite
common zeros will be $(N+1)^2 - 2N = N^2 + 1$, in agreement with previous
results \cite{witt90}.

Dehomogenizing the polynomials at these points, we find
\begin{eqnarray}
P_1^{\bar{Z}=1} & = & (z-wu) \prod_i (1 - u\bar{z_i}) - u^2 
\sum_i m_i \prod_{j \neq i} (1 - u\bar{z}_j),  \\ 
P_2^{\bar{Z}=1} & = & (1 - \bar{w}u) \prod_i (z - uz_i) - u^2 
\sum_i m_i \prod_{j \neq i} (z - uz_j),  \\ 
P_1^{Z=1} & = & (1 - wu) \prod_i (\bar{z} - u\bar{z}_i) - u^2
\sum_i m_i \prod_{j \neq i} (\bar{z} - u\bar{z}_j), \\ 
P_2^{Z=1} & = & (\bar{z} - \bar{w}u) \prod_i (1 - uz_i) - u^2
\sum_i m_i \prod_{j \neq i} (1 - uz_j).
\end{eqnarray}
Because of the complex conjugation symmetry of these expressions,
it suffices to examine the branches of, say, $P_i^{\bar{Z}=1}$ to
deduce the others.
The Newton diagrams of these are shown in Figure 3.
In each case, the Puiseux series are ordinary power series. 
For $P_1$ there is a single branch $u = z/w + \cdots$, while for
$P_2$ there are $N$ branches $u = z/z_i - (m_i / z_i^3) z^2 + \cdots$.

\begin{figure}
\begin{picture}(440,240)(0,-10)
\put(0,0){\vector(0,1){240}}
\put(0,0){\vector(1,0){120}}
\put(0,40){\circle*{5}}
\put(0,80){\circle*{5}}
\put(0,120){\circle*{5}}
\put(0,160){\circle*{5}}
\put(0,200){\circle*{5}}
\put(40,0){\circle*{5}}
\put(40,40){\circle*{5}}
\put(40,80){\circle*{5}}
\put(40,120){\circle*{5}}
\put(40,160){\circle*{5}}
\put(0,40){\line(1,-1){40}}
\put(0,246){\makebox(0,0){$u$}}
\put(126,0){\makebox(0,0){$z$}}
\put(-15,40){\makebox(0,0){$1$}}
\put(-15,160){\makebox(0,0){$N$}}
\put(-30,200){\makebox(0,0){$N+1$}}
\put(40,-15){\makebox(0,0){$1$}}

\put(200,0){\vector(0,1){240}}
\put(200,0){\vector(1,0){240}}
\put(200,200){\circle*{5}}
\put(200,160){\circle*{5}}
\put(240,160){\circle*{5}}
\put(240,120){\circle*{5}}
\put(280,120){\circle*{5}}
\put(280,80){\circle*{5}}
\put(320,80){\circle*{5}}
\put(320,40){\circle*{5}}
\put(360,40){\circle*{5}}
\put(360,0){\circle*{5}}
\put(200,160){\line(1,-1){160}}
\put(200,246){\makebox(0,0){$u$}}
\put(446,0){\makebox(0,0){$z$}}
\put(197,40){\line(1,0){6}}
\put(185,40){\makebox(0,0){$1$}}
\put(185,160){\makebox(0,0){$N$}}
\put(170,200){\makebox(0,0){$N+1$}}
\put(240,-3){\line(0,1){6}}
\put(240,-15){\makebox(0,0){$1$}}
\put(360,-15){\makebox(0,0){$N$}}

\end{picture}
\caption{The Newton diagrams for $P_1^{\bar{Z}=1}$ (left)
and $P_2^{\bar{Z}=1}$ (right) for $N$ point masses.
The case $N=4$ is shown.}
\end{figure}
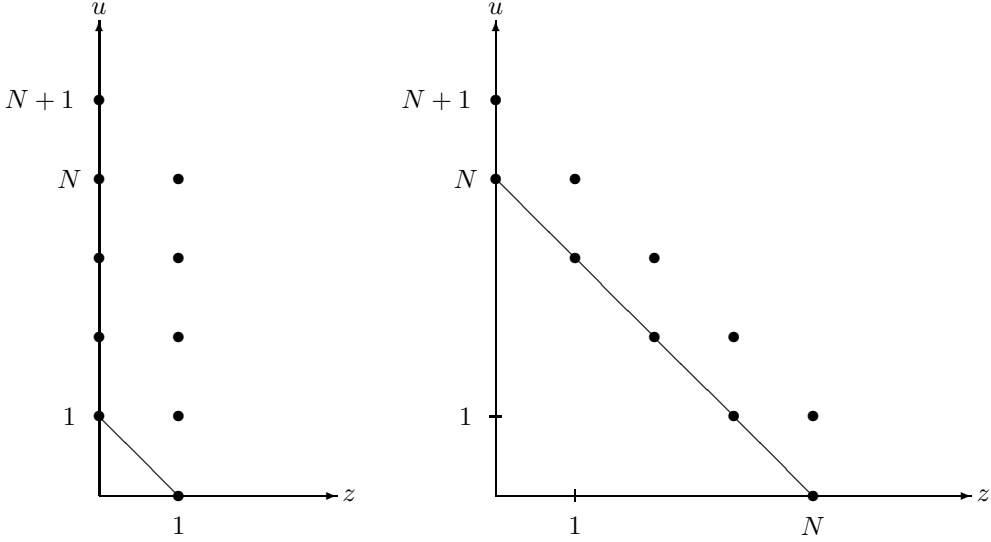

It is now straightforward to compute the residues
\begin{equation}
{\rm Res~} \frac{Z^k U^{2-k} \prod_i (Z - z_i U) (\bar{Z} - \bar{z}_i U)
\,d(Z/U)\,d(\bar{Z}/U) }{P_1^h P_2^h}
\end{equation}
and determine the moments.
The first two are $M = \Tr \mu = 1,$ and 
$\Tr z\mu = w + \sum_i \frac{m_i}{\bar{w} - \bar{z}_i}$.
The single branch contributes $w^k$ to the $k$th moment, while the 
contribution of the $N$ branches becomes progressively more complicated.

We have also considered a generalization of the model to include 
external shear.  This brings the lens equations to the form\cite{witt90}
\begin{equation}
z - w - \gamma \bar{z} - \sum_{i=1}^N \frac{m_i}{\bar{z}-\bar{z}_i} = 0
\end{equation}
and its complex conjugate.
Defining $P_i^h$ as before, their zeros at infinity are
$[1,0,0]$ and $[\gamma,1,0]$ for $P_1^h$, and 
$[0,1,0]$ and $[1,\gamma,0]$ for $P_2^h$.
For $\gamma \neq 0,1$ there are no common zeros at infinity.
Factoring the denominator of $\omega^h$ as $(U^n P_1^h) (P_2^h)$, 
the residue theorem gives the sum over the finite common roots as minus
the sum of residues at the infinite roots of $P_2^h$.
We find, for example, the total magnification $1/(1 - \gamma^2)$, and the
first moment 
\begin{equation}
\Tr z\mu = \frac{w + \gamma \bar{w}}{(1 - \gamma^2)^2}.
\end{equation}
Note that the first moment's dependence upon the lens' positions
has vanished.

\begin{table}
\begin{tabular}{lccc}
\toprule
Model & $\psi$ & $\sum_{i}\mu_i$ & $\sum_{i}\mu_i z_i$\\
\colrule
point masses & $\sum_l m_l\log|{\vec\Theta}-{\vec\Theta_l}|$ & 1 & 
$z_s+\sum_{l}{{m_l}\over{{\bar z_s}-{\bar z_l}}}$\\
point masses + shear & $\psi_{\rm pm}+{\gamma\over 2}r^2\cos 2\theta$
& $1/(1-\gamma^2)$ & ${1\over{(1-\gamma^2)^2}}(z_s+\gamma{\bar z_s})$\\
SIE & $bR$ & 2 & $2z_s$\\
SIS + elliptical & $br+\gamma br\cos 2 \theta$ & 1 & 
%$(x_s(1+2\gamma)-{{s^3\cos(3\theta_s)}\over{32b^2\gamma^2}},
%y_s(1-2\gamma)+{{s^3\sin(3\theta_s)}\over{32b^2\gamma^2}})$\\
$z_s + 2\gamma{\bar z_s} - {{{\bar z_s}^3}\over{32b^2\gamma^2}}$\\
SIS + shear& $br+{\gamma\over 2}r^2\cos 2\theta $ & 
${2/{(1-\gamma^2)}}$ & ${2\over{(1-\gamma^2)^2}}(z_s+\gamma{\bar z_s})$\\
SIE + shear& $bR+{\gamma\over 2}r^2\cos 2(\theta-\theta_\gamma)$ & 
${2/{(1-\gamma^2)}}$ & ${2\over{(1-\gamma^2)^2}}(z_s+\gamma 
e^{2i\theta_\gamma}{\bar z_s})$\\
\botrule
\end{tabular}
\caption{Model lens potentials and results.  
The $l^{\rm th}$ lens has mass $m_l$ and position 
${\vec\Theta_l}$.  For the galaxy potentials, with only one lens, we 
choose coordinates centered on the lens and oriented along the ellipticity
or shear axes.
Here, the image position ${\vec\Theta}=(x,y)=(r\cos\theta,r\sin\theta)$
and source position ${\vec\beta}=(x_s,y_s)=(s\cos\theta_s,
s\sin\theta_s)$.  We use the complex notation\cite{witt90} 
$z=x+iy$, and similarly for $z_s, z_l$.
Also, $R=\sqrt{x^2+y^2/q^2}$ 
is an elliptical radial coordinate,
$\gamma$ is the strength of the shear, and $q$ is the axis ratio.}
\label{tab1}
\end{table}

\section{Discussion}
\label{sec:discuss}

We have introduced in this paper a new framework for analyzing
gravitational lens models.  The use of residue integrals makes 
clear the origin of the magnification relations, and facilitates
their computation for a wide class of model potentials.  We have
also applied this method to a series of models, confirmed and
extended previous results, and provided new magnification
relations for several models.  Although multidimensional 
residue calculus may be unfamiliar to readers from astronomy,
one may follow a simple procedure to perform the necessary 
integrals.  Basically, the procedure is as follows :
\newcounter{lister}
\begin{list}{\arabic{lister}.}{\usecounter{lister}}
\item From the stationarity equations, construct two polynomials
$P_1,P_2$ that simultaneously vanish at (and only at) the image
positions.
\item Define the ``Jacobian'' $J_P=\det[\partial(P_1,P_2)/\partial(x,y)]$,
and define $g(x,y) = J_P(x,y)\mu(x,y)$ where $\mu$ is the magnification.
\item Change to homogeneous coordinates: $(x,y)\to(X,Y,U)$ with
$x=X/U, y=Y/U$ and homogenize the polynomials by multiplying each
by the factor $U^{{\rm deg}\ P}$.  Also, multiply $g$ by 
$U^{{\rm deg}\ P_1+{\rm deg}\ P_2}$.
\item From the homogenized $g,P_1,P_2$, construct the 2-form 
$(g/P_1P_2)d({X\over U})d({Y\over U})$, and factor the denominator
into two groups; usually, the grouping $P_1, P_2$ suffices.
(Henceforth $P_1,P_2$ shall refer to the two groups, not the original
polynomials.) If the denominator contains explicit factors of $U$ then
redefine one of the polynomials, say $P_2$, to contain these
additional factors.
\item For $U=0$, find the points $(X,Y)$ where $P_1,P_2$
simultaneously vanish; these are the roots at infinity.
\item Pick one of the polynomials -- say $P_1$ -- and determine
the behavior of $P_1=0$ in the vicinity of each common root. 
First define coordinates for the neighborhood of the root.  For example,
if the root is $Y=0,U=0$, then a good choice would be
$(X,Y,U)=(1,y,u)$.  
As discussed, there are in general multiple branches of $P_1=0$
meeting at the root at infinity, each parametrized by a Puiseux
series $y=\sum_ia_iu^{\alpha_i}$.  The $\alpha_i$'s can be determined
from the Newton diagram, and by substituting in the specified 
$\alpha_i$'s one may solve for the coefficients $a_i$.
Usually, only the first one or two terms in the series are necessary.
\item Construct the quantity
$${{g(u,a_1u^{\alpha_1}+a_2u^{\alpha_2}+...)}\over
{{{\partial P_1}\over{\partial y}}(u,a_1u^{\alpha_1}+...)
P_2(u,a_1u^{\alpha_1}+...)}}$$
and pick out the term $\sim u^{-1}$.  The coefficient of $u^{-1}$ is
the contribution for this branch; summing over all the branches gives 
the residue for each root.
\item Repeat this procedure for all the roots at infinity, and sum
their residues.  Negating this quantity gives the sum of the residues 
at finite poles (the images).
\end{list}

The above is of course just a rough outline; section \ref{sec:residue}
describes the method in full detail.
Following this procedure, the results listed in Table \ref{tab1}
can be reproduced with minimal effort.  As mentioned earlier, the
Newton diagram method of determining the branches near each pole
is effective only if the lens equations can be brought into
polynomial form, such as for these simple potentials.  For more
complicated models, introduction of additional variables may be
necessary to handle fractional powers.  Our method may not be
applicable to models involving transcendental functions, in particular
functions with essential singularities at infinity.

\textcite{dalal} and \textcite{wm99} have considered the applicability
of such magnification relations to real gravitational lenses.  For
galaxies, \textcite{dalal} has shown that these relations can be
an aid in fitting models to lensed objects, or can be used
to rule out models {\sl a priori}.  \textcite{wm99} have shown,
however, that reliance upon simple galaxy models can be misleading,
when applied to realistic galaxy potentials.  This limits the 
applicability of magnification relations to making statements about 
models, as opposed to statements about the lenses themselves.
For microlensing, however, there is no doubt about the accuracy
of the point mass approximation, and as such our derived magnification
relations may be considered exact.  \textcite{wm95} have already 
shown how the total signed magnification (``zeroth moment'') can be 
used for binary microlensing to set lower limits on the overall
unsigned magnification, useful for example for detecting source blending.  
Although the multiple images
of a microlensing event cannot as yet be resolved, precluding
the present-day experimental verification of our prediction regarding 
the first moment, we are hopeful that the future advent of 
space-based interferometers will allow our microlensing
formulae to be tested observationally.

\acknowledgments{The authors would like to thank 
Wyn Evans, Geza Gyuk, Eduard Looijenga, 
Peter Teichner, Adrian Wadsworth, Nolan Wallach, and John Wavrik for many
helpful discussions.  This work was supported in part by the
U.S. Dept. of Energy under grant DEFG0390ER40546.}

%File lensapp, appendix to paper.

\appendix
\section*{Appendix : The homology class of the torus $T$}

Here we explain why the torus $T$ defining the local residue at a
singular point, taken to be the origin, is homologous to 
the sum of the Leray tori constructed from the branches of {\it either}
of the polynomials $P_i(x,y)$, and how to determine the correct orientations.
We are grateful to Eduard Looijenga and Peter Teichner 
for explaining the topology to us.
See also section 2.2 of \textcite{D92}.

We are working locally in a closed ball $B$ around the origin, and we denote
by $X$ the zero locus $P_1P_2=0$ within $B$.
The various tori define homology classes in $H_2(B-X)$.
It is known that $X$ is topologically a cone with vertex the origin and base
$X \cap \partial B$, which is a linked collection of topological circles
\cite{BK86,D92}.
Denote the several branches of $X$ as $X_i$, and the Leray torus built
on a given branch as $\delta C_i$.
Let $\sigma_{ij}$ be a path running from the origin to $\partial B$
along $X_i$, and returning to the origin along branch $X_j$.
A subset of these paths forms a basis for the relative homology group 
$H_1(X,\partial B)$.
Furthermore, by Alexander duality \cite{AY83},
an element of $H_2(B-X)$ is uniquely determined by its
linking numbers with these paths (indeed, with those in a basis alone).
(The linking number $l(c^1,c^2)$ of a 1-cycle with a 2-cycle in a 
4-ball is the intersection number of $c^1$ with any
3-manifold having boundary $c^2$.)

The Leray tori have linking numbers 
\begin{equation}
l(\sigma_{ij},\delta C_k) = \delta_{ik} - \delta_{jk}.\tag{A1}
\end{equation}
Indeed, the intersection of $\sigma_{ij}$ with a solid torus bounded
by $\delta C_k$ is the intersection of $\sigma_{ij}$ with $C_k$, which
is one point if the outward segment of $\sigma_{ij}$ lies on branch $X_k$,
and one point (with opposite orientation) if the returning segment does.

The sum $\sum_{P_1} \delta C_k$ of Leray tori built on branches of $P_1=0$
therefore has linking number with $\sigma_{ij}$ equal to 
$+1$ if $X_i$ is a branch of $P_1$ and $X_j$ is a branch of $P_2$,
$-1$ if vice-versa,
and $0$ if $X_i$ and $X_j$ are branches of the same polynomial.
The sum $\sum_{P_2} \delta C_k$ has the negatives of these linking numbers and
therefore represents the same homology class but with opposite orientation.

It remains to show that $T$ has the linking numbers of 
$\sum_{P_1} \delta C_k$.
A solid torus bounded by $T$ is given by 
$\{ |P_1| \le \epsilon,\;|P_2|=\epsilon \}$.
This meets any branch $X_k$ of $P_1=0$ given by a Puiseux series $y=p(x)$
in the locus $|P_2(x,p(x))| = \epsilon$ on $X_k$.
This set is topologically a circle around the origin, having
intersection number $+1$ with a path radially outward from the origin.
The solid torus meets no branch of $P_2=0$.
Therefore, radially outward (inward) paths on any branch of $P_1$
contribute $+1$ ($-1$) to intersection numbers with this solid torus,
while paths on branches of $P_2$ contribute nothing.
This duplicates the linking numbers of $\sum_{P_1} \delta C_k$.

The orientation for $T$ used in this argument is indeed the standard one
prescribed in section \ref{sec:residue}.
Using the Leray residue formula to evaluate
\begin{equation}
\left( \frac{1}{2\pi i} \right)^2 \int_T \frac{dP_1 \, dP_2}{P_1P_2}
\tag{A2}
\end{equation}
produces a positive contribution from every branch of $P_1$.

\end{document}